\documentclass[aps,prd,preprint,twoside,tightenlines,showpacs,amsmath,amssymb,nofootinbib,floatfix,superscriptaddress]{revtex4}
\usepackage{graphicx}
\usepackage{amsfonts}
\usepackage{color}
\usepackage[colorlinks=true,urlcolor=blue,linkcolor=blue,citecolor=blue]{hyperref}
\usepackage[T1]{fontenc}
\usepackage{lmodern}

\begin{document}
\renewcommand{\theenumi}{(\alph{enumi})}

\title{\boldmath $B^0$ and $B^0_s$ decays into $J/\psi$ $f_0(980)$ and $J/\psi$ $f_0(500)$ and the nature of the scalar resonances }
\date{\today}
\author{W.H. Liang}
\affiliation{Department of Physics, Guangxi Normal University, Guilin 541004, China}
\affiliation{Departamento de
F\'{\i}sica Te\'orica and IFIC, Centro Mixto Universidad de
Valencia-CSIC Institutos de Investigaci\'on de Paterna, Aptdo.
22085, 46071 Valencia, Spain}
\author{E.~Oset}
\affiliation{Departamento de
F\'{\i}sica Te\'orica and IFIC, Centro Mixto Universidad de
Valencia-CSIC Institutos de Investigaci\'on de Paterna, Aptdo.
22085, 46071 Valencia, Spain}

\begin{abstract}
We describe the $B^0$ and $B^0_s$ decays into $J/\psi$ $f_0(500)$ and $J/\psi$ $f_0(980)$ by taking into account the dominant process for the weak decay of $B^0$ and $B^0_s$  into $J/\psi$ and a $q \bar q$ component. After hadronization of this  $q \bar q$ component into pairs of pseudoscalar mesons we obtain certain weights for the meson-meson components and allow them to interact among themselves. The final state interaction of the meson-meson components, described in terms of chiral unitary theory, gives rise to the $f_0(980)$ and $f_0(500)$ resonances and we can obtain the $\pi^+ \pi^- $ invariant mass distributions after the decay of the resonances, which allows us to compare directly to the experiments. We obtain ratios of  $J/\psi$ $f_0(980)$ and $J/\psi$ $f_0(500)$ for each of the $B$ decays in quantitative agreement with experiment, with the $f_0(980)$ clearly dominant in the $B^0_s$ decay and the $f_0(500)$ in the $B^0$ decay.
\end{abstract}

\maketitle

\section{Introduction}
The LHCb Collaboration measured the  $B^0_s$ decays into $J/\psi$ and $\pi^+ \pi^-$ and observed a pronounced peak for the $f_0(980)$ \cite{Aaij:2011fx}. Simultaneously the signal for the $f_0(500)$ was found very small or non-existent. The Belle Collaboration followed suit and reported similar results \cite{Li:2011pg}, providing absolute rates for the $f_0(980)$ production with a branching ratio of the order of $10^{-4}$.
Results of this order of magnitude have been predicted using light cone QCD sum rules under the factorization assumption \cite{Colangelo:2010bg}.
The CDF Collaboration corroborated these latter results in \cite{Aaltonen:2011nk}. Further confirmation was provided by the D0 Collaboration in \cite{Abazov:2011hv}. The LHCb Collaboration has brought much information into the topic and in \cite{LHCb:2012ae} results are provided for the $\bar B^0_s$ decay into $J/\psi$ $f_0(980)$ followed by the $\pi^+ \pi^-$ decays of the  $f_0(980)$. Once again the $f_0(980)$ production is seen clearly while no evident signal is seen for the $f_0(500)$. The interesting thing is that in the analogous decay of $\bar B^0$ into $J/\psi$ and $\pi^+ \pi^-$ \cite{Aaij:2013zpt} a signal is seen for the $f_0(500)$ production and only a very small fraction is observed for the $f_0(980)$ production, with a relative rate of about (1-10)\% with respect to that of the $f_0(500)$. Research has followed in the same collaboration and in \cite{Aaij:2014emv} the  $\bar B^0_s$ into $J/\psi$ and $\pi^+ \pi^-$ is investigated. Once again a clear peak is observed for  $f_0(980)$ production, while $f_0(500)$ production is not observed. The $\bar B^0$ into $J/\psi$ and $\pi^+ \pi^-$ is further investigated in \cite{Aaij:2014siy} with a clear contribution from the $f_0(500)$ and no signal for the $f_0(980)$.

    Such striking behavior has served the authors of \cite{Stone:2013eaa} to conduct a theoretical study in which they discuss the repercussions of assuming the $f_0(980)$ and $f_0(500)$ to be $q\bar q $ states or tetraquarks. On the other hand, the last fifteen years have witnessed a spectacular advance in our understanding of the low energy scalar mesons, made possible by the studies of chiral perturbation theory \cite{Gasser:1983yg,Bernard:1995dp} complemented by the requirement of exact unitarity in coupled channels \cite{Kaiser:1995eg,Oller:2000ma}. For the case of the scalar mesons it was found that the treatment of the interaction of pseudoscalar mesons using chiral unitary theory gave rise to the appearance of the $f_0(500)$ and $f_0(980)$ resonances in a natural way, leading to what is called dynamically generated resonances \cite{npa,ramonet,kaiser,markushin,juanito,rios}. These ideas have been largely and successfully tested in a variety of reactions where the $f_0(500)$ and $f_0(980)$ are produced. This is the case of \cite{osetli} for the $J/\psi$ decay into $p \bar p$ and a pair or mesons, where the $f_0(500)$, $f_0(980)$ and $a_0(980)$ production is well reproduced, the $\gamma \gamma \to \pi \pi$ reaction \cite{gamagama}, the $\phi \to \gamma \pi \pi$ reaction \cite{osetuge,Palomar:2003rb}, the $J/\psi \to \phi (\omega) \pi \pi$ where different signals for the $f_0(500)$, $f_0(980)$ are observed depending on the reaction \cite{Meissner:2000bc,chiangpalo,Lahde:2006wr,Hanhart:2007bd,Roca:2012cv} and many others.

      The works of  \cite{Meissner:2000bc,chiangpalo}, where explicit
consideration of final state interaction of the $\pi \pi, ~K \bar K$ is
taken into account, are also used in weak decays of $B$ mesons, concretely
the $B \to \pi \pi K $ decay \cite{robert,bruno}. Other related papers
are those of \cite{cheng,bruno2}, where the $B \to f_0(980) K$ reaction
is studied and the $f_0(980)$ is treated as a $q \bar q$ state, or the
one of \cite{bruno3} that studies the $B_s \to f_0(980) J/\psi$, where
the amplitude is parametrized and fitted to data. The $B \to f_0(980) K$
reaction is also studied in \cite{lucio}, again parametrizing the
amplitude and discussing about the $f_0(980)$ nature as a $q \bar q$ or
a tetraquark state.

      The idea that we exploit here is to take the dominant mechanism for the weak decay of the $B$'s into $J/\psi$ and a primary $q \bar q$ pair, which is $d \bar d$ for $B^0$ decay and $s \bar s$ for $B^0_s$ decay. After this, the $q \bar q$ pair is allowed to hadronize into a pair of pseudoscalar mesons and only the relative weights of the different pairs of mesons is needed. Once the production of these meson pairs has been achieved, they are allowed to interact, for what chiral unitary theory in coupled channels is used, and automatically the  $f_0(500)$, $f_0(980)$ resonances are produced. We are then able to evaluate ratios of these productions in the different decays studied and we find indeed a striking dominance of the  $f_0(500)$ in the $B^0$ decay and of the
$f_0(980)$ in the $B^0_s$ decay in a very good quantitative agreement with experiment.

 \section{Formalism}

\begin{figure}[t!]\centering
\includegraphics[height=3.0cm,keepaspectratio]{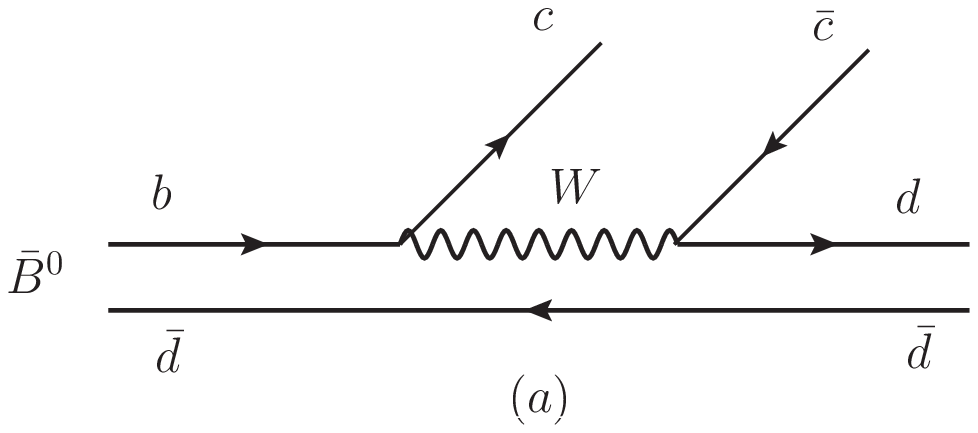}
\includegraphics[height=3.2cm,keepaspectratio]{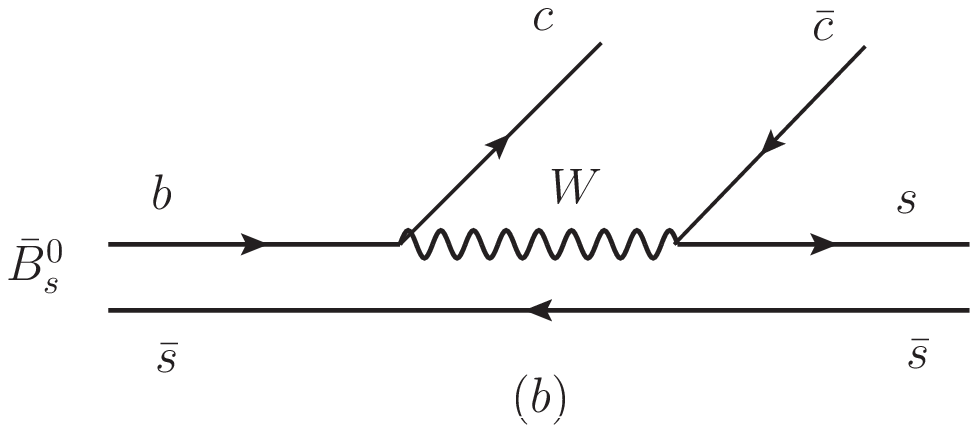}
\caption{Diagrams for the decay of $\bar B^0$ and $\bar B^0_s$ into $J/\psi$ and a primary $q\bar q$ pair, $d\bar d$ for $\bar B^0$ and $s\bar s$ for $\bar B^0_s$.
\label{fig:fig1}}
\end{figure}

Following \cite{Stone:2013eaa} and \cite{Aaij:2011fx,Li:2011pg,Aaltonen:2011nk,Abazov:2011hv,LHCb:2012ae,Aaij:2013zpt,
Aaij:2014emv,Aaij:2014siy} we take the dominant weak mechanism for $\bar{B}^0$ and $\bar{B}^0_s$ decays (the case is the same for $B^0$ and $B^0_s$ decays) which we depict in Fig.~\ref{fig:fig1}.

\begin{figure}[h!]\centering
\includegraphics[height=2.1cm,keepaspectratio]{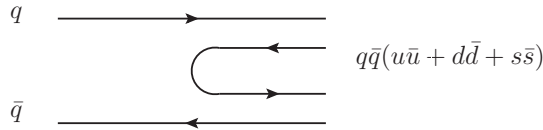}
\caption{Schematic representation of the hadronization $q\bar q \to q\bar q (u\bar u +d\bar d +s\bar s)$.\label{fig:fig2}}
\end{figure}

The differences between the two processes are: (i) $V_{cd}$ appears in the $Wcd$ vertex in $\bar B^0$ decay while $V_{cs}$ appears for the case of the $\bar B^0_s$ decay, where $V_{cd}$,$V_{cs}$ are the matrix elements of the Cabbibo-Kobayashi-Maskawa (CKM) matrix; (ii) One has a $d\bar d$ primary final hadron state in $\bar B^0$ decay and $s\bar s$ in $\bar B^0_s$ decay. Yet, one wishes to have $\pi^+ \pi^-$ in the final state as in the experiments. For this we need the hadronization. This is easily accomplished: schematically this process is as shown in Fig.~\ref{fig:fig2}, and adds an extra $q\bar q$ pair with the quantum numbers of the vacuum, $u\bar u +d\bar d +s\bar s$. Next step corresponds to writing the $q\bar q (u\bar u +d\bar d +s\bar s)$ combination in terms of pairs of mesons. For this purpose we follow the work of \cite{alberzou} and define the $q\bar q$ matrix $M$,
\begin{equation}\label{eq:1}
M=\left(
           \begin{array}{ccc}
             u\bar u & u \bar d & u\bar s \\
             d\bar u & d\bar d & d\bar s \\
             s\bar u & s\bar d & s\bar s \\
           \end{array}
         \right)
\end{equation}
which has the property
\begin{equation}\label{eq:2}
M\cdot M=M \times (u\bar u +d\bar d +s\bar s).
\end{equation}

Now, in terms of mesons the matrix $M$ corresponds to
\begin{equation}\label{eq:3}
\phi=\left(
           \begin{array}{ccc}
            \frac{1}{\sqrt{2}}\pi^0 +\frac{1}{\sqrt{6}} \eta & \pi^+  & K^+\\
             \pi^- &  -\frac{1}{\sqrt{2}}\pi^0 +\frac{1}{\sqrt{6}} \eta & K^0 \\
             K^- & \bar K^0 & -\frac{2}{\sqrt{6}} \eta \\
           \end{array}
         \right),
\end{equation}
where $\eta$ is actually $\eta_8$ (we shall come back to this). \footnote{In fact $M$ corresponds to $\phi + \frac{1}{\sqrt{3}} diag (\eta_1, \eta_1, \eta_1)$, where $\eta_1$ is the singlet of SU(3). In chiral perturbation theory the $\eta_1$ term is removed since it does not lead to any interaction. Here we should keep it to match the $M$ matrix, but, in as much as the $\eta_1$ is essentially the $\eta'$ state, since the mass of $\eta'$ is much bigger than any one of the octet, we can neglect the $\eta_1$ and work with the ordinary $\phi$ matrix.}  Hence, in terms of two pseudoscalars we have the correspondence:
\begin{align}\label{eq:4}
d\bar d (u\bar u +d\bar d +s\bar s) & \equiv  \left( \phi \cdot \phi \right)_{22}=\pi^- \pi^+ +\frac{1}{2}\pi^0 \pi^0
-\frac{1}{\sqrt{3}}\pi^0 \eta +K^0 \bar K^0 +\frac{1}{6}\eta \eta ,\nonumber \\
s\bar s (u\bar u +d\bar d +s\bar s) & \equiv  \left( \phi \cdot \phi \right)_{33}=K^- K^+ + K^0 \bar K^0 +\frac{4}{6}\eta \eta .
\end{align}

We can see that $\pi^+ \pi^-$ is only obtained in the first step in the $\bar B^0$ decay and not in $\bar B^0_s$ decay. However, upon rescattering of $K\bar K$ we also can get $\pi^+ \pi^-$ in the final state, as we shall see. Yet, knowing that the $f_0(980)$ couples strongly to $K\bar K$ and the $f_0(500)$ to $\pi \pi$, the meson-meson decomposition of Eqs.~(\ref{eq:4}) already tells us that the $\bar B^0$ decay will be dominated by $f_0(500)$ production and $\bar B^0_s$ decay by $f_0(980)$ production. The quantitative evaluation is done below.

\begin{figure}[t!]\centering
\includegraphics[height=3.0cm,keepaspectratio]{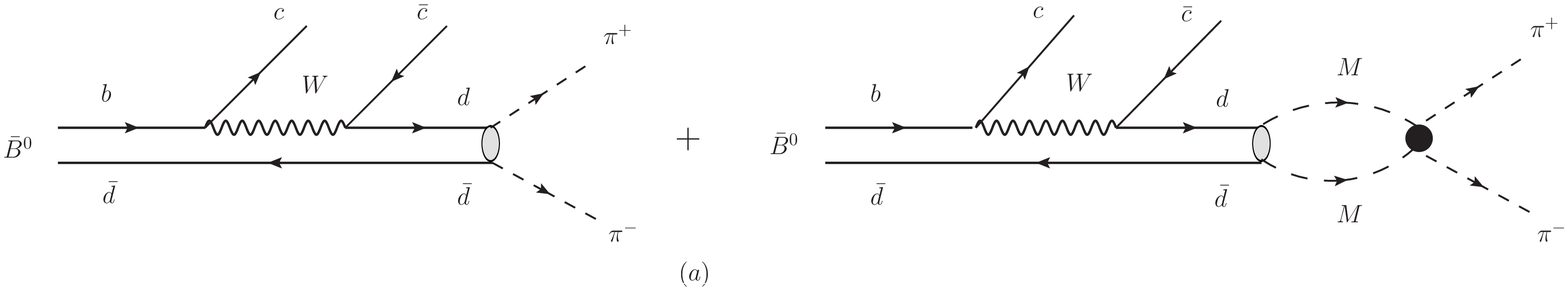}
\includegraphics[height=3.0cm,keepaspectratio]{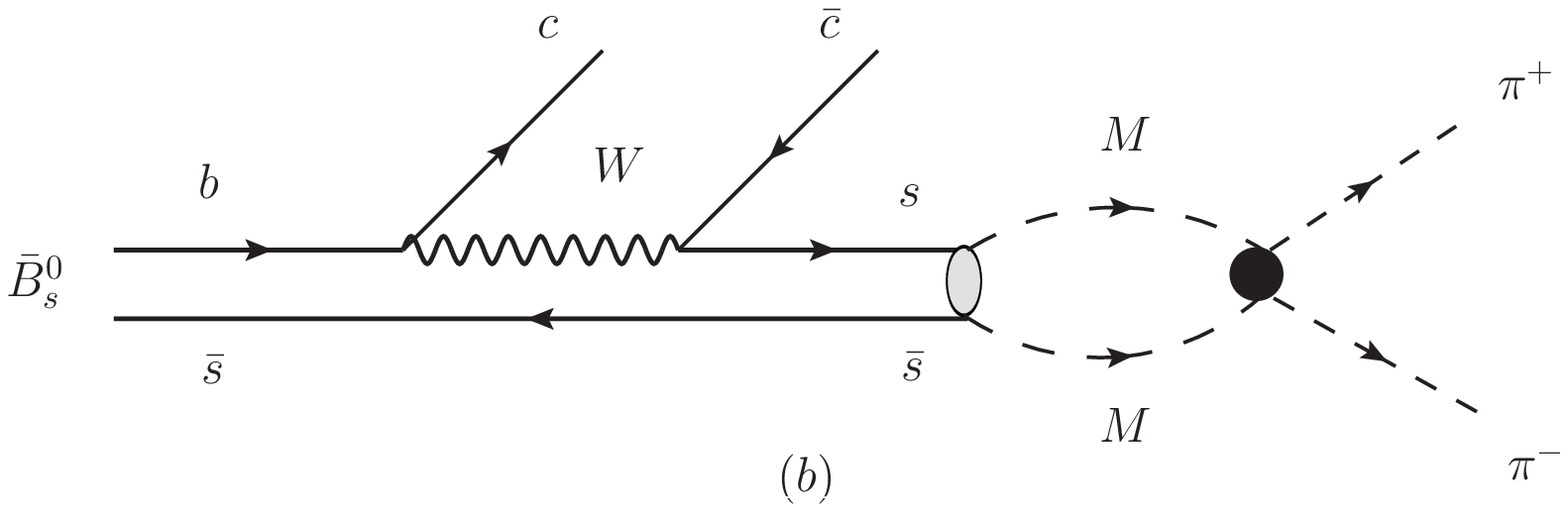}
\caption{Diagrammatic representation of $\pi^+ \pi^-$, via direct plus rescattering mechanisms in $\bar B^0$ decay (a), and via rescattering for $\bar B ^0_s$ decay (b).\label{fig:fig3}}
\end{figure}

Let us call $V_P$ the production vertex which contains all dynamical factors common to both reactions. The $\pi^+ \pi^-$ production will proceed via primary production or final state interaction as depicted in Fig.~\ref{fig:fig3}.

The amplitudes for $\pi^+ \pi^-$ production are given by
\begin{align}\label{eq:5}
t(\bar B^0 \to J/\psi \pi^+ \pi^-) & = V_P V_{cd}
(
1+G_{\pi^+ \pi^-} t_{\pi^+ \pi^- \to \pi^+ \pi^-} +\frac{1}{2} \frac{1}{2} G_{\pi^0 \pi^0} t_{\pi^0 \pi^0 \to \pi^+ \pi^-}  \nonumber \\
&~~ +G_{K^0 \bar K^0} t_{K^0 \bar K^0 \to \pi^+ \pi^-} +\frac{1}{6} \frac{1}{2} G_{\eta \eta} t_{\eta \eta \to \pi^+ \pi^-}) ,\nonumber \\
t(\bar B^0_s \to J/\psi \pi^+ \pi^-) & = V_P V_{cs} ( G_{K^+ K^-} t_{K^+ K^- \to \pi^+ \pi^-} +  G_{K^0 \bar K^0} t_{K^0 \bar K^0 \to \pi^+ \pi^-}  +\frac{4}{6} \frac{1}{2} G_{\eta \eta} t_{\eta \eta \to \pi^+ \pi^-} )~,
\end{align}
where $G_i$ are the loop functions of two meson propagators
\begin{equation}
G_i (s) = i \int\frac{d^{4}q}{(2\pi)^{4}}\frac{1}{(P-q)^{2}-m^2_1+i\varepsilon}\,\frac{1}{q^{2}-m^2_2+i\varepsilon},
\label{eq:G}
\end{equation}
where $m_1, ~m_2$ are the masses of the mesons in the $i$-channel, $q$ is the four-momentum of one meson, and $P$ is the total four-momentum of the system, thus, $s=P^2$.
The integral is performed integrating exactly for $q^0$ and implementing a cutoff $\Lambda$ of the order on $1~ \rm{GeV}/c$ for the three momentum.
The elements $t_{ij}$ are the scattering matrices for transitions of channel $i$ to $j$. According to \cite{npa} this matrix is given by
\begin{equation}\label{eq:BSeq}
t = [1-VG]^{-1} V,
\end{equation}
and the $V$ matrix is taken from \cite{npa} complemented with the matrix elements of the $\eta \eta$ channels, which we have taken from \cite{danydan}.
Numbering the channels as 1 for $\pi^+ \pi^-$, 2 for $\pi^0 \pi^0$, 3 for $K^+ K^-$, 4 for $K^0 \bar K^0$ and 5 for $\eta \eta$, the matrix elements projecting into $S$-wave are
\begin{align}\label{eq:Vkernel}
&V_{11}=-\frac{1}{2f^2}s,~&&V_{12}=-\frac{1}{\sqrt{2}f^2}(s-m_{\pi}^2),~&&V_{13}=-\frac{1}{4f^2}s,\nonumber \\
&V_{14}=-\frac{1}{4f^2}s,~&&V_{15}=-\frac{1}{3\sqrt{2}f^2}m_{\pi}^2,~&&V_{22}=-\frac{1}{2f^2}m_{\pi}^2,\nonumber \\
&V_{23}=-\frac{1}{4\sqrt{2}f^2}s,~&&V_{24}=-\frac{1}{4\sqrt{2}f^2}s,~&&V_{25}=-\frac{1}{6f^2}m_{\pi}^2, \\
&V_{33}=-\frac{1}{2f^2}s,~&&V_{34}=-\frac{1}{4f^2}s,~&&V_{35}=-\frac{1}{12\sqrt{2}f^2}(9s-6m_{\eta}^2-2m_{\pi}^2),\nonumber \\
&V_{44}=-\frac{1}{2f^2}s,~&&V_{45}=-\frac{1}{12\sqrt{2}f^2}(9s-6m_{\eta}^2-2m_{\pi}^2),~&&V_{55}=-\frac{1}{18f^2}(16m_{K}^2-7m_{\pi}^2), \nonumber
\end{align}
 where $f=93~{\rm MeV}$ is the pion decay constant and the unitary normalization $|\eta \eta\rangle \to \frac{1}{\sqrt{2}}|\eta \eta\rangle$, $|\pi^0 \pi^0\rangle \to \frac{1}{\sqrt{2}}|\pi^0 \pi^0\rangle$ has been taken to easily account for the identity of the particles when using the $G$ function without an extra factor in Eq.~(\ref{eq:BSeq}). The good normalization of the $t$ matrices must be restored in Eqs.~(\ref{eq:5}).

In \cite{npa} only the $\pi \pi, K\bar K$, channels were considered, while here, as in \cite{kaiser}, we also include the $\eta \eta$ channel. The threshold of this channel is far away from the $f_0(500)$ and not so much from the $f_0(980)$. The results obtained with or without the $\eta \eta$ channel are very similar but the cutoff needed with two channels is $\Lambda=903 ~{\rm MeV}/c$, while it is $600 ~{\rm MeV}/c$ for the three channels. The effective inclusion of new channels by means of a change in the cutoff is common place in coupled channels problems.

Another point to consider is that in Eqs.~(\ref{eq:Vkernel}) one is assuming the physical $\eta$ to be $\eta_8$ of SU(3), however, there is a small mixing with the singlet $\eta'$. One has the $\eta=\cos \theta_P ~\eta_8 -\sin \theta_P ~\eta_1$. Since the $\eta_1$ has null interaction with the chiral Lagrangians, one must multiply the matrix elements of Eqs.~(\ref{eq:Vkernel}) by $\cos \theta_P$ for each $\eta$ involved in the process. Since $\theta_P =-14.34^{\circ}$ \cite{Ambrosino:2009sc}, the effect of the mixing is very small but we take it into account.

In Eqs.~(\ref{eq:5}) we made use of the fact that both the  $f_0(500)$ and  $f_0(980)$ appear in relative $L=0$ meson-meson orbital angular momentum, and then $\pi \pi$ in the final state selects $I=0$, hence, the $\pi^0 \eta$ intermediate state does not contribute. Note also that with respect to the weights of the meson-meson components in Eqs.~(\ref{eq:4}) we have added a factor $1/2$ for the propagation of the $\pi^0 \pi^0$ and $\eta \eta$ states which involve identical particles.

The $V_{cd}$ and $V_{cs}$ CKM matrix elements are in this case related to the Cabbibo angle,
\begin{align}\label{eq:Cabbibo}
&V_{cd}=-\sin\theta_c =-0.22534,\nonumber \\
&V_{cs}=\cos\theta_c =0.97427.
\end{align}
One final element of information is needed to complete the formula for $d\Gamma/dM_{inv}$, with $M_{inv}$ the $\pi^+ \pi^-$ invariant mass, which is the fact that in a $0^- \to 1^- 0^+$ transition we shall need an $L'=1$ for the $J/\psi$ to match angular momentum conservation. Hence, $V_P = A~p_{J/\psi} \cos \theta$, and we assume $A$ to be constant (equal to $1$ in the calculations). Thus,
\begin{equation}\label{eq:dGamma}
  \frac{d \Gamma}{d M_{inv}}=\frac{1}{(2\pi)^3}\frac{1}{4M_{\bar B_j}^2}\frac{1}{3}p_{J/\psi}^2 p_{J/\psi} \tilde{p}_{\pi} {\overline{ \sum}} \sum \left| \tilde{t}_{\bar B^0_j \to J/\psi \pi^+ \pi^-} \right|^2,
\end{equation}
where the factor $1/3$ is coming from the integral of $\cos^2 \theta$ and $\tilde{t}_{\bar B^0_j \to J/\psi \pi^+ \pi^-}$ is $t_{\bar B^0_j \to J/\psi \pi^+ \pi^-}/(p_{J/\psi} \cos \theta)$, which depends on the $\pi^+ \pi^-$ invariant mass. In
Eq.~(\ref{eq:dGamma}) $p_{J/\psi}$ is the $J/\psi$ momentum in the global CM frame ($\bar B$ at rest) and $\tilde{p}_{\pi}$ is the pion momentum in the $\pi^+ \pi^-$ rest frame,
\begin{equation}\label{eq:pJpsi}
 p_{J/\psi}=\frac{\lambda^{1/2}(M_{\bar B}^2, M_{J/\psi}^2, M_{inv}^2)}{2M_{\bar B}},~~~~~~~\tilde{p}_{\pi}=\frac{\lambda^{1/2}(M_{inv}^2, m_{\pi}^2, m_{\pi}^2)}{2M_{inv}}.
\end{equation}

\section{Results}
\begin{figure}[ht!]\centering
\includegraphics[height=7.0cm,keepaspectratio]{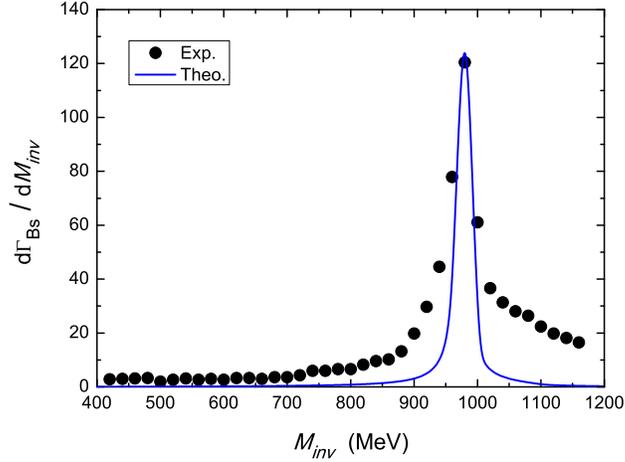}
\caption{$\pi^+ \pi^-$ invariant mass distribution for the $\bar B^0_s \to J/\psi \pi^+ \pi^-$ decay, with arbitrary normalization. Data from \cite{Aaij:2014emv}.  \label{fig:dGammaBs}}
\end{figure}

In Fig.~\ref{fig:dGammaBs} we show the $\pi^+ \pi^-$ invariant mass distribution for the case of the $\bar B^0_s \to J/\psi \pi^+ \pi^-$ decay.
We compare our results with the data of  \cite{Aaij:2014emv} where more statistics has been accumulated than in the earlier run of \cite{Aaij:2011fx}. The data are collected in bins of 20 MeV. Thus, we fold our mass distributions with the size of the bins, and
compare the results obtained with those in fig.~14 of \cite{Aaij:2014emv}. The data points, given in events per bin in the experimental paper, are normalized here to match the theoretical strength at the peak of the distribution. We can see that the agreement is quantitatively good. We observe an appreciable peak for $f_0(980)$ production and basically no trace for $f_0(500)$ production.
At lower invariant masses both the theory and experiment show a very small strength, with the theory below the data. The agreement is even better with the dashed line in fig.~14 of \cite{Aaij:2014emv} where a small background has been subtracted. At invariant masses above the $f_0(980)$ peak, contribution from higher energy resonances, which we do not consider, is expected \cite{Aaij:2014emv}.

The second of Eqs.~(\ref{eq:5}) tells us why the $f_0(500)$ contribution is so small. All intermediate states involved $K\bar K, \eta\eta$, have a mass in the $1~{\rm GeV}$ region and the $G$ functions are small at lower energies. Furthermore, the coupling of the $f_0(500)$ to both $K\bar K$ and $\eta \eta$ is also extremely small, such that the $t$ matrices involved have also small magnitudes.

\begin{figure}[ht!]\centering
\includegraphics[height=7.0cm,keepaspectratio]{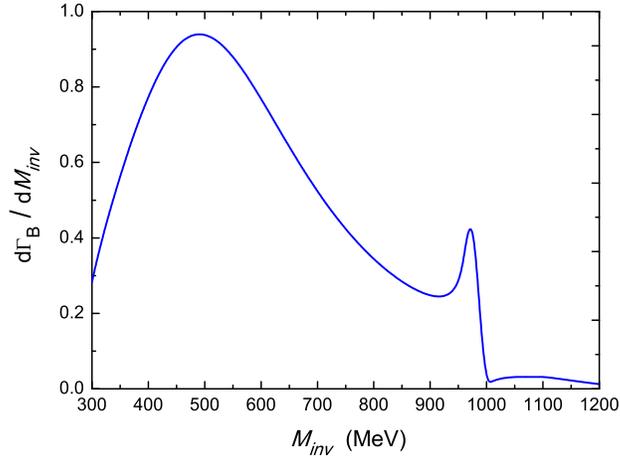}
\caption{$\pi^+ \pi^-$ invariant mass distribution for the $\bar B^0 \to J/\psi \pi^+ \pi^-$ decay, with arbitrary normalization.\label{fig:dGammaB}}
\end{figure}

Note that in this decay we could have also $J/\psi$ and vector meson production, but the $s\bar s$ component would give $\phi$ production which does not decay to $\pi \pi$. The case is quite different for the $\bar B^0 \to J/\psi \pi^+ \pi^-$ decay, because now we can also produce $J/\psi \rho$ ($\rho \to \pi^+ \pi^-$) decay and in fact this takes quite a large fraction of the $J/\psi \pi^+ \pi^-$ decay, as seen in \cite{Aaij:2014siy}. We plot our relative $S$-wave $\pi^+ \pi^-$ production for the $\bar B^0 \to J/\psi \pi^+ \pi^-$ decay in Fig.~\ref{fig:dGammaB}.

We can see that the $f_0(500)$ production is clearly dominant. The $f_0(980)$ shows up as a small peak. The strength of the peak of the $f_0(980)$ is about 1/3 of that of the $f_0(500)$. A direct comparison with the extracted results for the $f_0(500)$ and $f_0(980)$ from the partial wave analysis of \cite{Aaij:2014siy} (see fig. 13 of \cite{Aaij:2014siy}) is not possible because of the smallness of the $f_0(980)$ signal, which is of the order of the fluctuations of the data and is not shown in the figure. However, a test can be done to compare the results: If we integrate the strength of the two resonances over the invariant mass distribution we find
\begin{equation}\label{eq:ratio}
  \frac{{\cal B}[\bar B^0 \to J/\psi f_0(980), f_0(980)\to \pi^+ \pi^-]}{{\cal B}[\bar B^0 \to J/\psi f_0(500), f_0(500)\to \pi^+ \pi^-]} =0.033\pm 0.007,
\end{equation}
with an admitted 20\% uncertainty from the decomposition of the strength in Fig.~\ref{fig:dGammaB} into the two resonances. The most recent experimental result is
\begin{equation}\label{eq:LHCb2014}
  \left( 0.6^{+0.7+3.3}_{-0.4-2.6} \right) \times 10^{-2}
\end{equation}
from \cite{Aaij:2014siy} which superseded the earlier one
\begin{equation}\label{eq:LHCb2013}
  \left( 9.5^{+6.7}_{-3.4}\pm 3 \right) \times 10^{-2}
\end{equation}
from \cite{Aaij:2013zpt}.
The central value that we obtain is five times bigger than the central value of the experiment in Eq.~(\ref{eq:LHCb2014}), yet, by considering the errors in Eq.~(\ref{eq:LHCb2014}) we get a band for the experiment of $0 \sim 0.046$ and our results are within this band.

There is another point in our calculations worth noting. The normalization of  Figs.~\ref{fig:dGammaBs} and ~\ref{fig:dGammaB} is arbitrary but the relative size is what the theory predicts. It is easy to compute that
\begin{equation}
\frac{\Gamma(B^0\to J/\psi f_0(500))}{\Gamma(B^0_s \to J/\psi f_0(980))}\simeq \left( 4.5 \pm 1.0 \right)\times 10^{-2}.
\label{eq:ratio-th}
\end{equation}
This number is in agreement within errors with the band of $(2.08 \sim 4.13)\times 10^{-2}$ that one obtains from the branching fractions of $9.60^{+3.79}_{-1.20} \times 10^{-6}$ for $\bar B^0 \to J/\psi f_0(500)$ from \cite{Aaij:2013zpt} and $3.40^{+0.63}_{-0.16} \times 10^{-4}$ for $\bar B^0_s \to J/\psi f_0(980)$ from \cite{LHCb:2012ae} .

Added to the results obtained for many other reactions, as quoted in the Introduction, the present reactions come to give extra support to the idea originated from chiral unitary theory that the $f_0(500)$ and $f_0(980)$ resonances are dynamically generated from the interaction of pseudoscalar mesons and could be interpreted as a kind of molecular states of meson-meson with the largest component $\pi \pi$ for the $f_0(500)$ and $K\bar K$ for the $f_0(980)$.

So far we have assumed that $V_P$ is constant up to the $P$-wave factor. Actually there is a form factor for the transition that depends on the momentum transfer. Then it could be different for $f_0(500)$ or $f_0(980)$ production. Yet, in \cite{liform} this form factor is evaluated and it is found that $F^{\sigma}_{B^0_s}(m^2_{J/\psi})/F^{f_0}_{B^0_s}(m^2_{J/\psi})=1$, where $\sigma, f_0$ stand for the $f_0(500), f_0(980)$. In \cite{ochsform} the same results are assumed, as well as in \cite{Stone:2013eaa}, where by analogy $F^{\sigma}_{B^0}(m^2_{J/\psi})/F^{f_0}_{B^0}(m^2_{J/\psi})$ is also assumed to be unity. In addition, in \cite{Stone:2013eaa} it is also found from analysis of the experiment that $F^{f_0}_{B^0_s}(m^2_{J/\psi})/F^{\sigma}_{B^0}(m^2_{J/\psi})$ is compatible with unity.  These findings justify the assumption made of a constant form factor in the range of energies discussed here.

\section{conclusion}

  In this paper we have addressed the $B^0$ and $B^0_s$ decays into $J/\psi$ $f_0(980)$ and $J/\psi$ $f_0(500)$ and have looked at the $f_0(980)$ and $f_0(500)$ production in both cases. The formalism is easy to follow: starting from the dominant weak decay process we have $J/\psi ~ d \bar d$ production for the $B^0$ decay and $J/\psi ~ s \bar s$ production in the $B^0_s$ decay. Upon hadronization into meson-meson components we realize that the $d \bar d$ contains $\pi \pi$ components while the $s \bar s$ does not and contains mostly $K \bar K$. This already hints at the dominance of $f_0(500)$ in the $B^0$ decay and the $f_0(980)$ in the  $B^0_s$ decay since the $f_0(500)$ couples mostly to $\pi \pi$, while the $f_0(980)$ couples mostly to $K \bar K$. We have made a quantitative study by allowing the primary produced meson-meson pairs  to interact among themselves, using for this the chiral unitary approach, and then the two resonances are generated. We observe a pronounced $f_0(980)$ peak in the $\pi \pi$ invariant mass distribution for the $B^0_s$ decay and no visible trace of the  $f_0(500)$ production, like in the experiment.  On the other hand, in the $B^0_s$ decay, both the $f_0(500)$ and $f_0(980)$ excitations are visible, but the $f_0(980)$ production represents only a very small fraction of the $f_0(500)$, also in quantitative agreement with experiment.

    The results obtained here add to the analysis of other reactions where both the $f_0(500)$ and $f_0(980)$ are also produced. The systematic and accurate agreement of the predictions of chiral unitary theory with experiment is remarkable and gives a strong support to the idea of the low lying scalar mesons as being formed from the interaction of pairs of pseudoscalar mesons, qualifying as dynamically generated resonances.

\section*{Acknowledgments}

We would like to thank Diego Milanes for useful discussions and motivation to do this work, Luis Roca for checking part of the calculations and Melahat Bayar for checking the calculations and detecting a bug in the program. We also thank A. Ramos for pointing up the difference of the matrices $M$ and $\phi$ for pseudoscalars and S. Stone and L. Zhang for enlightening discussions concerning the experiments. This work is partly supported by the Spanish Ministerio de Economia
y Competitividad and European FEDER funds under the contract number
FIS2011-28853-C02-01 and FIS2011-28853-C02-02, and the Generalitat
Valenciana in the program Prometeo, 2009/090. We acknowledge the
support of the European Community-Research Infrastructure
Integrating Activity Study of Strongly Interacting Matter (acronym
HadronPhysics3, Grant Agreement n. 283286) under the Seventh
Framework Programme of EU. This work is also partly supported by the
National Natural Science Foundation of China under Grant No. 11165005.

\bibliographystyle{plain}

\end{document}